\documentclass[proof]{WileyASNA-v1}

\articletype{Article Type}%

\received{26 April 2016}
\revised{6 June 2016}
\accepted{6 June 2016}

\newcommand{\ergs}{\mbox{ erg s}^{-1}}

\newcommand{\kms}{\mbox{ km s}^{-1}}
\newcommand{\kpc}{\mbox{ kpc}}
\newcommand{\pc}{\mbox{ pc}}
\newcommand{\Msun}{M_\odot}
\newcommand{\mach}{\mathcal{M}}
\newcommand{\vir}{\alpha_{\rm vir}}
\newcommand{\sfrff}{\mbox{SFR}_{\rm ff}}
\newcommand{\scrit}{s_{\rm \small crit}}
\newcommand{\eq}[1]{Eq.~(\ref{#1})}

\begin{document}

\title{Resolved simulations of jet-ISM interaction: Implications for gas dynamics and star formation }

\author[1]{Dipanjan Mukherjee*}
\author[2]{Geoffrey V. Bicknell}
\author[3]{Alexander Y. Wagner}

\authormark{Mukherjee \textsc{et al}}

\address[1]{Inter-University Centre for Astronomy and Astrophysics (IUCAA), Pune University, Ganeshkhind, Pune - 411007, India}
\address[2]{Research School of Astronomy and Astrophysics, Australian National University, Canberra, ACT 2611, Australia}
\address[3]{Center for Computational Sciences, University of Tsukuba, 1-1-1 Tennodai, Tsukuba, Ibaraki, 305-8577}

\corres{ *\email{dipanjan@iucaa.in}}

\abstract{  
Relativistic jets can interact with the ambient gas distribution of the host galaxy, before breaking out to larger scales. In the past decade several studies have simulated jet-driven outflows to understand how they affect the nearby environment, and over what spatial and temporal scales such interactions occur. The simulations are able to capture the interaction of the jets with the turbulent clumpy interstellar medium and the resultant energetics of the gas. In this review we summarise the results of such recent studies and discuss their implications on the evolution of the dynamics of the gas distribution and the star formation rate. 

}


\maketitle

\jnlcitation{\cname{
\author{D. Mukherjee}, 
\author{G.V. Bicknell}, and  
\author{A.Y. Wagner}}, (\cyear{2021}),
\ctitle{Resolved simulations jet-ISM interaction: Implications for gas dynamics and star formation}, \cjournal{Astron. Nachr.}, \cvol{01}.}

\section{Introduction}\label{sec.intro}
Feedback from active galactic nuclei (AGN) in the form of outflows is an important driver of galaxy evolution \citep{silk98a,fabian12a}. Large scale relativistic jets powered by the central black hole are one such means by which the AGN can transfer energy to its environment, affecting the stellar mass assembly of galaxies \citep{fabian12a}. Although relativistic jets have primarily been considered in the context of large scale heating of the circum-galactic environment to prevent cooling flows \citep{gaspari11a,fabian12a,yang16a}, studies have established that jets can also strongly interact with the gas in the host galaxy itself, before evolving to larger structures. 

Understanding the impact of compact jets evolving through their host is also of interest for the studies of a class of radio galaxies, called the Giga-Hertz Peaked Spectrum (GPS) or Compact Steep Spectrum (CSS) sources \citep{odea98a,odea21a}. These sources are characterised by a turnover in their radio spectrum, resulting either from synchrotron self-absorption inside the radio emitting plasma itself \citep{kellerman69a} or free-free absorption by the ionised dense gas, external to the jet \citep{bicknell97a,bicknell18a}. Which of the above processes dominate is still a question of debate, with both giving equally acceptable interpretation of observations \citep{odea21a}. However, recently, several radio galaxies with steeply inverted optically-thick spectral indices $> 2.5$ have been discovered \citep{gopal14a,callingham15a,mhaskey19a,mhaskey19b}. This supports a free-free process as the absorbing mechanism and implies the presence of dense gas surrounding the jets. Additionally, high rotation measures and lower polarisation in such sources \citep{odea98a,odea21a}, along with spatially resolved observations of shocked emission from the local gas \citep{zovaro19a,zovaro19b}, further support the scenario of such sources harbouring jets that are evolving through a gas-rich environment \citep[see also][]{patil20a}. Thus the physics of jet-ISM interactions is an important topic to understand in the context of young evolving jets.

The inability of large scale cosmological simulations to resolve the interaction of AGN driven outflows with individual dense molecular clouds due to restricted spatial resolutions, calls for zoomed-in simulations of isolated galaxies. That such an interaction between a fast supersonic jet and a thermal cloud may occur, has long been theorised \citep{blandford79b}, and further promoted by the `dentist drill' scenario of evolving young jets \citep{scheuer82a,bicknell97a}.  With the advent of better computational resources, non-relativistic simulations of jet-cloud interactions have been performed by a number of authors \citep[such as][]{bicknell03a,saxton05a,sutherland07a,gaibler11a,gaibler12a,dugan17a,cielo18a}. In recent times, development of robust numerical techniques to solve for the relativistic hydrodynamic flows have resulted in several papers exploring how relativistic jets break out of a dense ISM of its host \citep[e.g.][]{wagner11a,mukherjee16a}. We review below some of the results of these works and their implications on the evolution of the host galaxy.

\section{Resolved simulations of Jet-ISM interaction}
Interaction of a relativistic jet with an ISM with fractal clouds were investigated in a series of papers, starting with \citet{wagner11a} and \citet{wagner12a}. A relativistic jet was propagated through a lognormally distributed static cloud distribution placed into an otherwise homogeneous atmosphere. These were followed up by simulations where the fractal ISM with a turbulent velocity dispersion was first allowed to settle in a galactic potential followed by a jet launched from the centre \citep{mukherjee16a,mukherjee17a,bicknell18a,mukherjee18a,mukherjee18b}. Some of the key results of these studies can be summarised as follows:
\begin{itemize}
\item \textbf{The flood and channel phase:}
\begin{figure*}
	\centering
	\includegraphics[width = 15 cm, keepaspectratio]{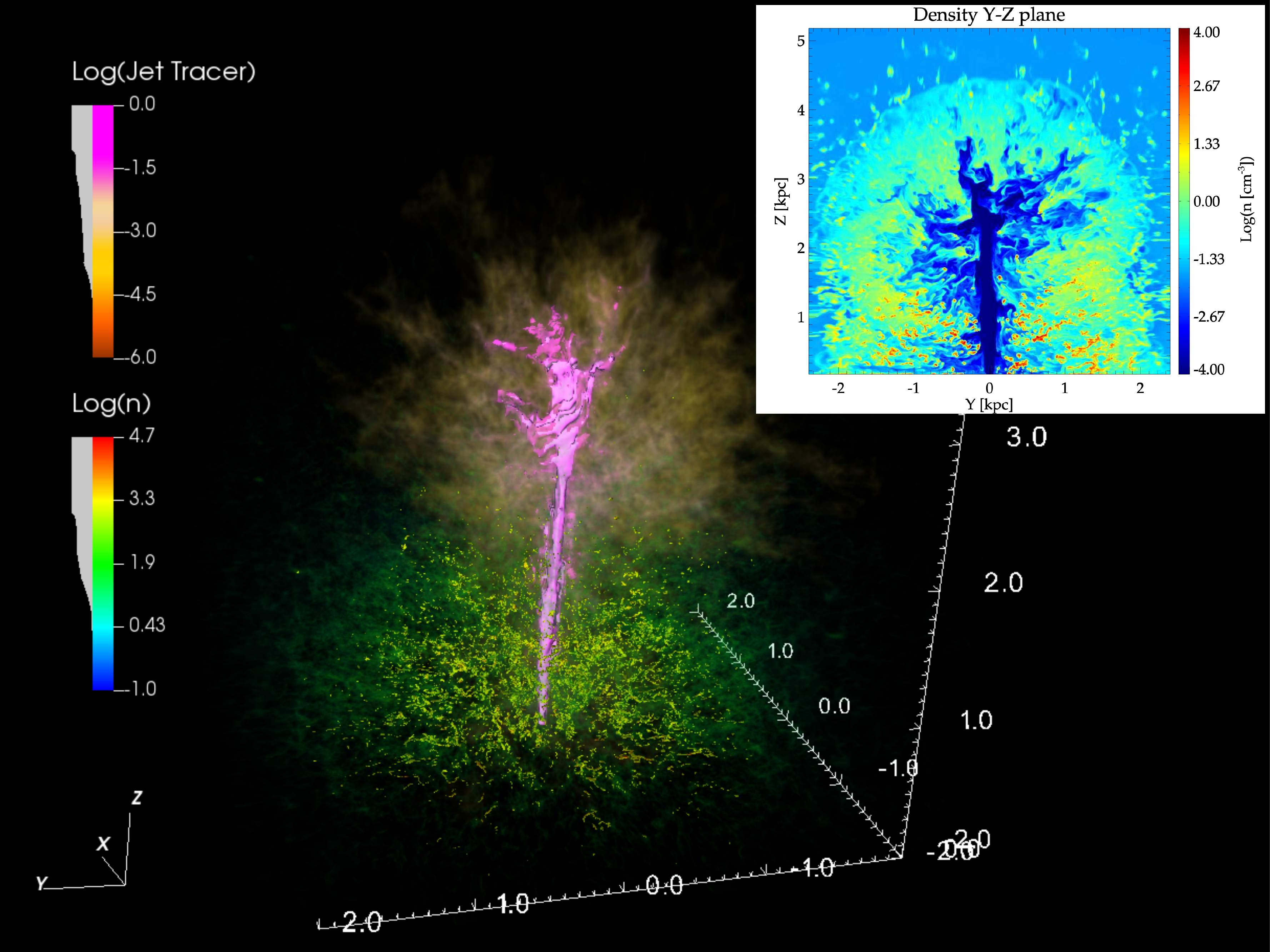}
        \caption{3D Volume rendering of the jet tracer and dense clouds of simulation E of \citet{bicknell18a} at $t\sim 1$ Myr. The primary axis of the jet with the highest concentration of the tracer is shown in magenta in the centre. The head of the jet stream shows bifurcations into filaments due to interaction with clouds. This is better represented in the image of the density in the $y_Z$ plane shown in the inset in the top right corner. Surrounding the central jet axis is a diffuse filamentary cloud of jet plasma, depicted in orange, forming the so-called `flood and channel' phase. This results from the jet interacting with the clouds as it breaks out from central clouds shown in yellow-green colors. The active interaction of the jet with clouds in the iSM is also evident from the cross-section of the density in the $Y-Z$ plane.}
	\label{fig.floodchannel}
\end{figure*}
The jets pass through a `flood and channel' evolutionary phase \citep{sutherland07a,wagner11a} as they drill through the clouds before breakout. During such a phase, the jet streams get deflected and broken into smaller channels by the clouds. This spreads the collimated jet into filamentary low density streams of non-thermal plasma, percolating through the gaps of the thermal clouds and engulfing them. A representation of this is depicted in Fig.~\ref{fig.floodchannel}, which shows the volume rendering of the jet tracer\footnote{The 'jet-tracer' is a passive scalar advected with the fluid flow. It denotes the location of the jet plasma injected into the computational domain.} of simulation E from \citet{bicknell18a}. Here a jet of power $P_{\rm jet} = 10^{45} \ergs$ is ploughing through a cloud distribution extended up to $\sim 2$ kpc from the centre. The main jet is a thin collimated structure (in magenta), which shows some bifurcation at its head due to active interaction with clouds dragged by the jet flow. However, it is surrounded by filaments of low density  ($\sim 10^{-3}$ of the main jet) jet tracer material.

The extended distribution of the low density jet plasma implies that the jet can potentially drive shocks into the otherwise cold ISM, up to distances a few kpc away from the main jet axis. Observing such tenuous extended jet plasma may be difficult, as their surface brightness is nearly $\sim3-4$ orders of magnitude lower than the peak value at the jet-head \citep{bicknell18a}. However, signatures of shocks from such  material will be evident from multi-phase observations of the thermal gas. This has been recently demonstrated in spatially resolved observations of jet-ISM interactions in some galaxies \citet{zovaro19a,zovaro19b}, where the extent of the shocked gas was found to be several times larger ($\sim 1-2$ kpc), than the apparent size of the radio jet ($\lesssim 100$ pc).

\item \textbf{Impact of resolving cloud structures:}
An important aspect of these results was the strong dependence of the mean outflow velocities of the clouds on the cloud size and ISM filling factor \citep{wagner12a}. An ISM with a lower filling factor or smaller cloud sizes results in larger mean outflow velocities and hence more gas removal, similar to the results from large scale cosmological simulations. However, larger cloud sizes or a higher filling factor of the ISM may result in lower outflow speeds. This is more drastic in the simulations with an external gravitational field \citep{mukherjee16a,mukherjee17a} where the net mass-weighted mean radial velocity is negative for most simulations. This follows from the fact that for the same power ($P_w$) of an incident wind, the approximate velocity of the cloud of density $\rho_c$ will be: $v_c \sim \left(P_w/ \rho_c \right)^{1/2}$. Thus, low density, unresolved cloud structures will be more easily swept out of the galaxy, as often found in large scale simulations \citep{springel05a}, whereas resolved simulations with larger density contrasts show more resistance to outflows.

\item \textbf{Galactic fountain:}
The jets can transfer a significant fraction of their energy and momentum to the clouds, implying strong coupling with the ISM. The simulations of \citet{wagner11a} and \citet{wagner12a} found the mechanical advantage (ratio of cloud momentum to that of the jet) to be greater than unity. A significant fraction of the jet's energy ($\lesssim 30\%$) can be imparted to the clouds as kinetic energy. However, this does not imply that the AGN driven outflows can clear the gas from the galaxy and hence substantially reduce the star forming fuel, as envisaged earlier in large scale simulations \citep{springel05a}. 

Although the jets can cause outflows of diffuse ionised gas exceeding speeds of $\sim 1000 \kms$, the net mass weighted radial velocity of the entire ISM is not higher than the escape velocity \citep{mukherjee16a,mukherjee17a}. Considering a purely ballistic trajectory of the clouds, only a small fraction of the dense gas ($\lesssim 1 - 10 \%$) would reach a height of $\sim 10$ kpc. This proposes a `galactic fountain' scenario, where local outflows triggered by the jet remain within the galaxy's potential and will eventually fall back in. Thus, although the jets may not be very efficient in clearing out the gas in the galaxy, they do increase the kinetic energy budget of the dense gas, which will induce turbulence \citep[as also shown in][]{mukherjee18b}. 

\end{itemize}

\section{Impact on star formation of the host}
How AGN driven outflows may influence the stellar mass assembly of the host galaxy is an important question. In this context, relativistic jets have primarily been considered for their role in the large scale heating of the circum-galactic medium and offsetting cooling flows \citep{gaspari12a,fabian12a,yang16a}. However, there has been growing evidence both from theoretical simulations, as detailed above, as well as observations (both direct and indirect) that jets can both enhance and quench the star formation rate (SFR) of their host. We summarise below the current understanding on how these processes may be mediated, and the future work needed to unravel them better.

\subsection{Induced SFR: positive feedback}
Jets aligned with star forming regions have often been considered to be evidence of jet induced star formation e.g. 4C 41.17 \citep{bicknell00a,nesvadba20a}, Minkowskis object \citep{salome15a,lacy17a}, 3C 285 \citep{salome15a}. Shocks from relativistic jets can be more efficient in compressing the ambient gas than non-relativistic outflows. For example, the advance speed of a relativistic shock extending into a medium is \citep[see eq.~(19) - eq.~(22) of ][ for the detailed derivation]{mukherjee20a}:
\begin{equation} 
v_{c} \approx \gamma_j v_j \left(\frac{\rho_j}{\rho_c}\right)^{1/2} \left\lbrack 1 + \frac{\Gamma p_j}{(\Gamma -1) \rho_j c^2}\right\rbrack^{1/2} \propto \gamma_j  v_{c, \mbox{\small NR}}.
\end{equation}
Here ($\rho_j,v_j$) are the density and velocity of the jet or wind and ($\rho_c,v_{c, \mbox{\small NR}}$) are the pre-shock (up-stream) density in the cloud and the velocity of the shock in the cloud. The corresponding non-relativistic velocity is given by  $v_{c, \mbox{\small NR}} =  v_j \left(\frac{\rho_j}{\rho_c}\right)^{1/2}$ \citep{fragile04a}. Thus a relativistic outflow can create shock speeds higher by the Lorentz factor of the jet. Also, if the jets are collimated, they can retain their momentum up to much larger distances, as opposed to adiabatically expanding energy driven winds, whose advance speeds are likely to decrease with radius as $V \propto t^{-(2-\beta)/(5-\beta)}$ \citep[see Appendix A of][]{mukherjee18a}.

This leads to an overall enhancement of the density of the gas in the simulations of jet-ISM interactions \citep{sutherland07a,wagner16a,mukherjee16a,mukherjee17a}. Such strong compression are expected to enhance gravitational collapse and hence aid the formation of stars. This has been investigated by several authors, as compression from direct interactions of jets with the ISM \citep{fragile04a,gaibler12a,fragile17a,bourne17a,mukherjee18a} or by indirect compression of the extended gas disk due to the over-pressured lobes driven by the jets \citep{bieri16a}.

\subsection{Turbulence induced negative feedback}
However, an often important point missed in these analyses is the impact of turbulence induced by the jet. Resolved simulations of \citet{mukherjee18a} have indeed shown that the turbulent velocity dispersion increases by $\sim 2-4$ times its initial value (depending on jet power). This is mediated both by direct impact of the jet as well as shocks at the outer edges of the disk due to the backflow of the jet material, forming a high pressure bubble that engulfs the disk. Several observational studies \citep{nesvadba10a,nesvadba11a,alatalo11a} have found the star formation rate to be significantly reduced in host galaxies with a radio jet of modest power. The outflows launched in such systems are not expected to significantly deplete the dense gas. 

\begin{figure}
	\centering
	\includegraphics[width = 9 cm, keepaspectratio]{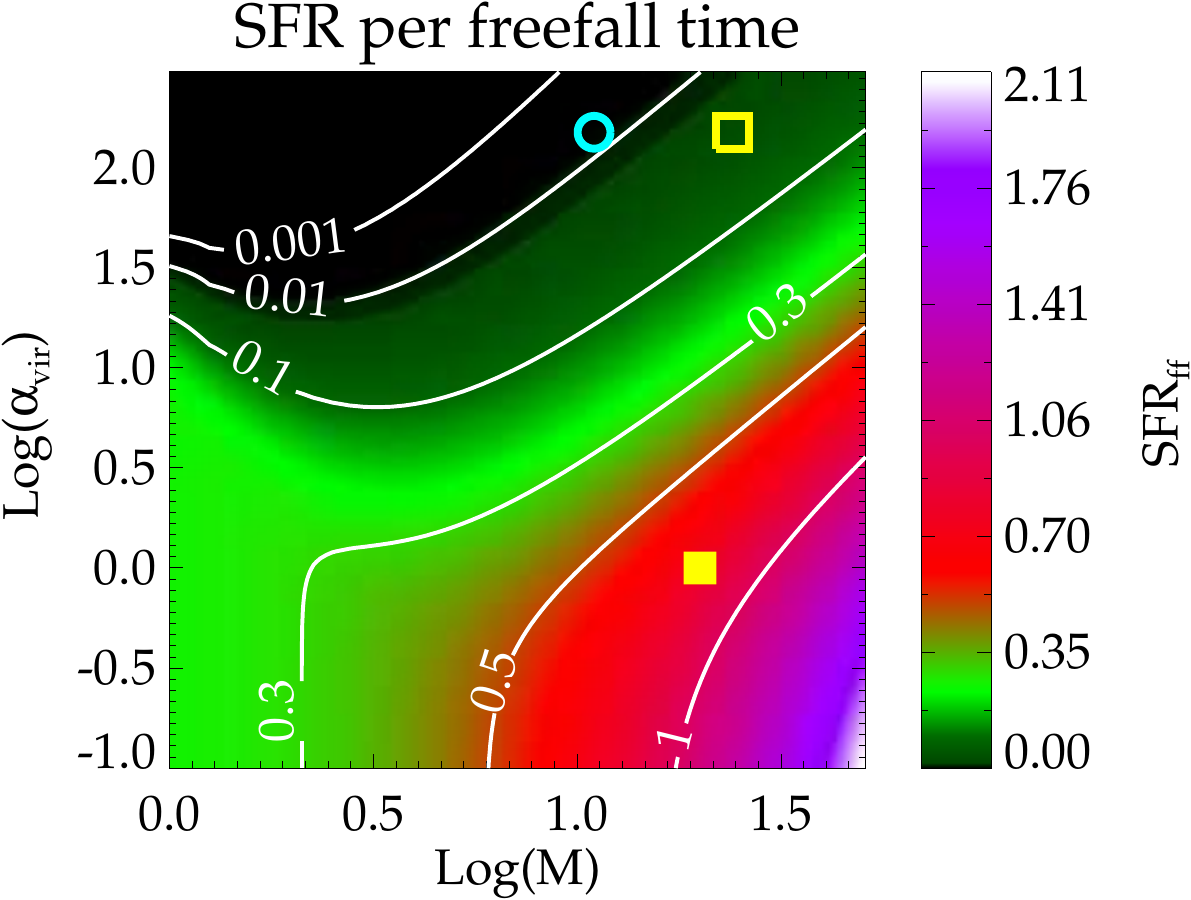}
        \caption{ Distribution of the $\sfrff$ given in \eq{eq.sfrff} as a function of the virial parameter ($\vir$) and mach number ($\mach$). The $\sfrff$ of the three different cases in Table~\ref{tab.sfrff} are marked as: i) Case A in filled yellow square, ii) Case B in open yellow square and iii) Case C in open cyan circle. }
	\label{fig.sfrff}
\end{figure}
The theory of turbulence regulated star formation \citep{krumholz05a,padoan11a,hennebelle11a,federrath12a} indicates that besides the density, the two key parameters that affect the SFR are: a) the virial parameter ($\vir = 2 E_{\rm kin}/E_{\rm grav}$), which is the ratio of twice the kinetic energy and gravitational potential energy, b) the rms mach number ($\mach$) of the turbulent gas. Using the above, one can compute the star formation rate per freefall time \citep[hereafter $\sfrff$, see][]{krumholz05a}. The total star formation rate of the cloud will then be $\mbox{SFR} = (M/t_{\rm ff}) \sfrff$. 

The $\sfrff$ for a multi-freefall model of turbulence regulated star formation is given by \citep{federrath12a}
\begin{align}\label{eq.sfrff}
&\sfrff = \frac{\epsilon}{2 \phi_t}\exp\left(\frac{3}{8} \sigma_s^2\right) \left\lbrack 1 + \mbox{erf}\left(\frac{\sigma_s^2 - s_{\rm crit}}{\sqrt{2 \sigma_s^2}}\right)\right\rbrack, \\
&\sigma_s^2 = \ln(1 + b^2 \mach^2), \, \scrit = \ln\left\lbrack \left(\frac{\pi^2}{5}\right) \phi_x^2 \vir \mach^2 \right\rbrack \nonumber.
\end{align}
The density of the cloud is assumed to be represented by a lognormal probability distribution function (PDF), expressed in terms of $s = \ln(\rho/\rho_0)$, $\rho_0$ being the mean density. $\sigma_s$ denotes the dispersion of the density PDF. We refer the reader to \citet{krumholz05a} and \citet{federrath12a} for detailed derivation of \eq{eq.sfrff}. In Fig.~\ref{fig.sfrff}, we have plotted \eq{eq.sfrff} for different values of $\vir$ and $\mach$. We have assumed the best fit values of the  various parameters, as found in \citet{federrath12a}: $\phi_x = 0.19$, $b =0.4$, $1/\phi_t = 0.49$ and $\epsilon = 0.5$

\begin{table}
\caption{$\sfrff$ for different states of a spherical gas cloud of size $L=20\pc$ and mass $M=10^5 M_\odot$}\label{tab.sfrff}
\centering
\begin{tabular}{| l | c | c | c | c | c |}
\hline
Case 	       &$\sigma_V$ ($\kms$) & $\vir$ & $T$ (K) & $\mach$  & $\sfrff$       \\
\hline
A              & 5                  & 1      & 10      & 20       & 0.77      \\   
B              & 61                 & 150    & 100     & 24.4     & 0.03      \\   
C              & 61                 & 150    & 500     & 11       & 0.006      \\ 
\hline  
\end{tabular} 
\end{table}
Using \eq{eq.sfrff}, one can demonstrate quantitatively the impact of increasing the turbulent velocity dispersion on the star formation rate. Let us a consider a cloud of mass $M\sim 10^5 \Msun$, with a velocity dispersion $\sigma_V \simeq 5 \kms$ and size $\sim 20 \pc$, typical of nearby molecular clouds in the Milky Way \citep{miville17a,solomon87a}. Let us approximate the cloud to be a homogeneous sphere for analytical ease, although far from a realistic description of typical filamentary molecular cloud.  The virial parameter for such a cloud will be \citep{federrath12a}
\begin{align}
\alpha_{\rm vir} &\approx 0.96 \times \left(\frac{\sigma_V}{5 \kms}\right)^2 \left(\frac{L}{20 \pc}\right)\left(\frac{M}{10^5 M_\odot}\right)^{-1} \label{eq.vir}
\end{align}
Assuming a temperature of 10 K and mean molecular weight of $\mu = 2.3$ \citep{pineda08a,federrath12a}, gives a sound speed of $c_s \simeq 0.25 \kms (T/10 \mbox{K})^{1/2}$ and a mach number $\mach \simeq 20 $. The above values result in a $\sfrff \sim 0.77$ (Case A of Table~\ref{tab.sfrff}), placing the cloud in the lower left corner of Fig.~\ref{fig.sfrff}, denoted by the filled yellow square.

Let us now consider that the above cloud lies in an ISM impacted by a jet, which induces internal turbulence with a line of sight dispersion of $\sigma_{1D} \sim 250 \kms$, similar to observed values in 3C 326 \citep{nesvadba11a} and IC 5063 \citep{morganti15a}. This would imply an internal 3D dispersion of $\sigma_{3D} = \sqrt{3} \sigma_{1D} \simeq 433 \kms$.  The observed values are for gas at kpc scales. Assuming the ISM to be an extended turbulent distribution, the velocity dispersion will approximately scale with length as \citep{larson81a,solomon87a,federrath12a}
\begin{equation}
\sigma_{V, 20\pc} = 250 \kms \left(\frac{20 \pc}{1 \kpc}\right)^{1/2} \simeq 61 \kms
\end{equation}
This would imply a virial parameter (from \eq{eq.vir}) of $\vir \approx 150$. The shocks induced by the jet will also heat up the gas, increasing its sound speed. The temperature of such a turbulent gas can be a few hundred Kelvins, as inferred from excitation temperature of molecular gas \citep[e.g. $T\sim100$ K by][]{dasyra16a}, or the presence of shocked H2 gas in several systems \citep[e.g. in 4C 31.04][]{zovaro19a}. We have listed in Table~\ref{tab.sfrff} the $\sfrff$ for two values of the temperature of the jet disturbed gas, and also demarcated their positions in Fig.~\ref{fig.sfrff}. As can be seen, the resultant $\sfrff$ is reduced by more than an order of magnitude.

The above analysis, though very simplistic and approximate, gives some indications that the SFR in the jet disturbed ISM can indeed be suppressed by turbulence, as is predicted in several systems \citep{nesvadba10a,nesvadba11a,alatalo11a,alatalo15a}. Which of the above feedback mechanisms (positive vs negative) dominates in a system is a complex problem and requires applying the above prescription of turbulence regulated star formation to simulations of jet-ISM interaction \citep[as done in][]{mandal21a}.

\section{Summary}
Relativistic jets from the central supermassive black holes can significantly affect the gas distribution of the host galaxy. Although jets can potentially couple strongly with the gas, the resultant velocities may not be sufficient to unbind a significant fraction of the gas. However, the jet driven outflows can potentially affect the star formation rate (SFR) of the host in two ways. They can either compress gas by direct interaction, enhancing the SFR as a positive feedback. Alternatively, the turbulence induced by the jets can significantly lower the star formation rate due to support from the turbulent velocity dispersion and increased temperature of the gas (and hence lower Mach number). To further test these implications, more resolved simulations of jet-ISM interactions, with better subgrid models of star formation, is required in future.

\section*{Acknowledgments}
The results discussed here spans the works from several publications in the past few years and ongoing works. The authors would like to thank Ralph Sutherland, Joseph Silk, Nicole Nesvadba, Raffaella Morganti, Christoph Federrath, Ankush Mandal and M. Meenakshi for insightful interactions; and the anonymous referee for helpful suggestions.

\def\apj{ApJ}%
\def\mnras{MNRAS}%
\def\aap{A\&A}%
\def\apjl{ApJ}
\def\physrep{PhR}
\def\apjs{ApJS}
\def\pasa{PASA}
\def\pasj{PASJ}
\def\nat{Nature}
\def\memsai{MmSAI}
\def\aj{AJ}%
\def\aaps{A\&AS}%
\def\iaucirc{IAU~Circ.}%
\def\sovast{Soviet~Ast.}%
\def\apss{Ap\&SS}

\bibliography{dmrefs}

\section*{Author Biography}
\begin{boxtext}
\textbf{Dipanjan Mukherjee} Assistant professor at the Inter-University Centre for Astronomy and Astrophysics (IUCAA), Pune, India. His interest lies in  simulating astrophysical flows, with a focus on AGN feedback, relativistic jets and related high energy phenomena.  
\end{boxtext}

\end{document}